# Barriers to Adopting Design for Assembly in Modular Product Architecture: Development of a Conceptual Model Through Content Analysis

**Fabio Marco Monetti, first author**[1]
Department of Production Engineering, KTH Royal Institute of Technology
Brinellvägen 68, 114 28, Stockholm
Sweden (SE)
monetti@kth.se

**Adam Lundström, second author**
Sweden Modular Management AB
Kungsgatan 56, 111 22 Stockholm
Sweden (SE)
adam.lundstrom@modularmanagement.com

**Antonio Maffei, third author**
Department of Production Engineering, KTH Royal Institute of Technology
Brinellvägen 68, 114 28, Stockholm
Sweden (SE)
maffei@kth.se

## ABSTRACT

*This study investigates the barriers to integrating Design for Assembly (DFA) principles within modular product architectures established using the Modular Function Deployment (MFD) method—a critical stage for deploying mass customization production while reducing costs. Despite the potential benefits of DFA, its application in modular architectures development remains underutilized, due to a mix of challenges. Through content analysis of qualitative data gathered from a focus group and interviews with industry experts and practitioners, we identified four major categories of such challenges, or barriers to adoption of DFA: technological, economic, regulatory, and organizational (TERO). Key challenges include compliance with regulatory requirements for data usage, intellectual property concerns, and limited availability of quantitative data in the initial stages of MFD.*
*The findings reveal that multidisciplinary collaboration is essential to addressing these barriers, as it enhances informed decision-making and eases the practical integration of DFA. By analyzing insights from both academic literature and industrial practice, this research develops a conceptual model that describes the main issues of applying DFA in MFD, providing a valuable guide for companies aiming to improve their modular products' assembly process. Ultimately, this study provides groundwork to support industry practitioners in overcoming existing barriers, promoting more cost-effective, high-quality modular design processes with the inclusion of efficient assembly considerations.*

## INTRODUCTION

Production systems nowadays must adapt to ever-changing product designs and customer demands to stay competitive. The growing complexity and variability of the market require the efficient reconfiguration of production systems and schedules to accommodate such changes [1]. Consequently, there is a shift from traditional rigid production systems to reconfigurable systems capable of adapting to evolving requirements, integrating new technologies, and optimizing resource utilization [2].

---

[1] Mail: monetti@kth.se; tel.: +46 73 461 89 25.

High-level reconfigurability requires a solid foundation in product design, with modularity playing a key role. Modular product architectures influence the product structure and the production system, supply chain, and reinforces flexibility, ultimately enabling a more customizable offering to the end user [3]. Additionally, in response to the growing trend of personalization, the range of options within modular product architectures continues to expand [4]. A structured approach to modular design reinforces scalability and customization in development and manufacturing processes, simultaneously reducing costs associated with production, storage, and inventory management, and improving the overall quality of the manufactured goods [5].

Many modularization methods can be found in the literature [6], along with related product platform development approaches [7]. Existing methods for creating modular product architectures, such as Modular Function Deployment (MFD) – based on research performed at KTH Royal Institute of Technology in the 1990s and the primary focus of this research – center on defining the architecture's building blocks, i.e., modules and interfaces, often neglecting crucial manufacturing and assembly factors [8].

The economic implications of different modularization strategies are not yet fully understood. It is though understood that neglecting assembly considerations can lead to higher production costs, inefficiencies, and challenges in achieving high-quality production, especially when there are numerous product variants [9, 10]. For example, if process costs are not factored in, a design with a higher product cost may be disregarded, even though the total cost, considering both product and process, could be lower due to reduced process costs [11].

For modularity to reach its full strategic potential, a streamlined and carefully planned assembly process is crucial [12]. Developing standardized, interchangeable interfaces should be complemented by minimizing the overall number of components and defining assembly operations, sequences, and strategies early in the process. This approach aligns directly with the goals of Design for Assembly (DFA), which aims to simplify assembly processes, reduce costs, and improve efficiency by streamlining design choices that facilitate easier and faster product assembly. Such integration enables manufacturing systems to adapt to changing market demands and production requirements [13].

Despite these advantages, the integration of DFA is often marginal in the conceptual phase of design, due to various challenges, including overlooking its potential for defining optimal manufacturing concepts [14]. These challenges were initially explored from an academic perspective in a previous study [15]. However, investigating the industrial perspective to understand how these challenges are perceived and addressed in production environments is essential. This research thus seeks to identify these barriers by gathering insights from industry experts through interviews and a focus group.

The objectives of this paper are to examine current practices in integrating DFA into modular product design and to identify potential barriers. By exploring the perspectives of practitioners and experts, this study aims to build on existing knowledge and develop a conceptual model offering practical recommendations for improving modularity design processes, specifically through the MFD method.

Drawing on secondary data from previous literature review [15] and other academic sources, as well as primary data gathered through interviews and a focus group, this paper identifies key themes that hinder the integration of DFA concepts in modular architectures created via MFD. A thorough content analysis process supports the inductive reasoning used to explore the phenomenon under investigation. This research provides a structured foundation for advancing modular product design, ultimately supporting the development of cost-effective and high-quality modular products and production systems.

The remainder of this article is structured as follows. The THEORETICAL BACKGROUND section introduces key principles and literature analysis, while the METHODOLOGY section details the research process. The RESULTS section present the content analysis outcomes, followed by the DISCUSSION of the findings, limitations and future work. Finally, the CONCLUSIONS section wraps up the paper.

## THEORETICAL BACKGROUND

This section introduces two key concepts central to the research: DFA and product modularity. Focusing on the MFD method, predominantly used in Sweden, where this research originated. It outlines the importance of considering assemblability in product design. The previously published literature review [15] that motivated this research also provided the main topics of discussion for qualitative data collection. Contributions from July 2022 to July 2024 were included to complement the earlier review endeavor.

## Design for Assembly

DFA concepts are crucial in product design, focusing on ease of assembly to reduce labor costs, minimize errors, and enhance product quality. Integrating DFA streamlines assembly processes, standardizes parts, and optimizes assembly directions [16]. DFA is well-established, tracing back to publications since the 1980s, such as Boothroyd and Dewhurst [17], Hitachi [18], and Lucas [19]. Since the turn of the century, automatic assembly has also gained attention with Design for Automatic Assembly (DFAA) [16].

While implementing DFA techniques can increase initial design costs and time, these are offset by substantial savings in later product lifecycle stages, including manufacturing, supply chain, logistics, and storage [8, 15]. The broader objectives include improving product quality, promoting inter-department communication and concurrent engineering, and fostering a sustainable manufacturing system [21]. These goals should be targeted from the outset of development [22]. Simplifying assembly by reducing part count and standardizing components significantly lowers manufacturing costs, labor, and material use, while also reducing defects and training requirements [9], yet early-stage DFA tools are underutilized, representing only 6 % of the total number of documented industry cases [20].

It is also worth mentioning 'assembly in the small' and 'assembly in the large'. The former deals with the detailed assembly of individual parts into a functioning product, focusing on technical solutions and component interactions. The latter takes a broader perspective, considering the economic, business, and institutional issues involved in product development. It looks at the architecture, defines the physical relationships between the product's constituents and relates them to the product's functions [21].

## Modular Function Deployment

Defining an appropriate product architecture is crucial for various aspects, from product development to managing variety. Companies can ensure ease of assembly, cost-effectiveness in production, and alignment with market requirements. Furthermore, a well-conceived product architecture promotes knowledge transfer and collaboration among departments and stakeholders [3].

Research shows that modular architecture rather than integral is better suited for achieving high product variety by dividing the product into independent modules [3]. It enhances product differentiation [23], supports flexible manufacturing, improves supplier cooperation [24] and allows for the creation of product families with a broader range of variants [25]. Additionally, modularity facilitates the development and automation of assembly systems, as each module is part of a distinct assembly subsystem [8].

To design modular products effectively, standard interfaces and generic modules that enable component reuse and interchangeability are crucial [26]. The process should address product functionalities and the interfaces between components [27].

In Sweden, MFD is widely employed to achieve these goals and help companies create modularity through a structured approach. MFD, developed in the 1990s at KTH, is an approach that supports the development of product modules based on customer values, product properties, and other critical factors. While MFD is instrumental in establishing architectures, it has certain limitations when it comes to DFA. It primarily focuses on breaking down products into modules but often lacks DFA integration. These considerations are addressed towards the end of the design cycle, rather than being incorporated from the outset. This delay can lead to suboptimal assembly, in terms of how modules are put together and the overall production system strategy. MFD does not inherently account for the ease of assembly until the whole architecture has been established, and then only suggests applying DFA to each individual module. Consequently, there is a missed opportunity to design modular products that are straightforward and cost-effective to assemble. Given these limitations, there is a pressing need for a holistic and integrated approach to modular product design that includes DFA from the conceptual phase.

## *Additional literature*

A previous literature review [15] revealed key contributions to understanding the state of the art in DFA integration within modularity development processes. This review informed the focus group discussion and interview topics, guiding the research.

Most existing contributions focus on product concepts or small-scale product families, typically aiming to improve the design quality of individual parts and streamline specific assembly tasks. However, there is a growing need for an expanded model that addresses assembly challenges at both system and organizational levels. Such a model would consider factors like assembly sequences, line layout, and production planning, all of which influence strategic decisions within companies. A standardized, structured methodology could help tailor assembly processes to align with unique company settings, environmental constraints, and operational demands, thereby supporting high-level management decisions that shape company strategy.

The literature also emphasizes the need for advanced software tools to enhance data collection, visualization, and knowledge sharing among stakeholders. Additionally, recent research has underscored the importance of integrating sustainability metrics, life cycle assessments, and circular economy principles to align with global sustainability initiatives and prioritize environmental considerations.

Key contributions from 2022 to 2024 were collected to enhance the review and provide targeted insights for actionable research. Findings from recent literature were used to identify specific areas for further investigation, with some examples provided in the following paragraphs.

Concepts of DFX strategies were presented, including a framework focused on environmental impact and production efficiency to ease knowledge acquisition and transfer in product development. This approach improved knowledge management and sharing across fields, supporting more effective design decisions [28]. Another study proposed an integrated approach using DFX guidelines to design circular and durable products, emphasizing reuse, repair, refurbishing, remanufacturing, and upgrading parts and products as essential components of sustainable design [29]. Additional research explored design for disassembly (DfD) methodologies and their applications, noting a gap in linking DfD with circular economy assessments. One study introduced a butterfly diagram that connects DfD parameters to circular economy indicators [30], highlighting the need to integrate DfD methodologies to enhance product sustainability [31].

A methodology for optimizing the design of a product family considering detailed CAD models of individual products was introduced. It aimed to achieve optimal standardization without compromising performance by using surrogate models to mitigate computational complexity [32]. A fully automated framework substituted the traditional sequential development process with a more efficient combined procedure, using regression models based on Gaussian process and artificial neural networks to achieve optimal design [33].

Optimization techniques are enabling faster and more effective design processes, essential for managing the complexities of product development. In this context, interdisciplinary knowledge is increasingly critical. The integration of artificial intelligence (AI) into the design process is gaining momentum, with AI-driven models and optimization methods proving valuable for handling complex design tasks more efficiently.

To complement the theoretical findings, we engaged with practitioners who are experts in the field. Using topic clusters from the literature review, we evaluated their perspectives to build a comprehensive understanding that integrates both theoretical insights and practical experience.

## METHODOLOGY

The paper followed a mixed methods approach to gather multiple sources of qualitative data, as outlined in Table 1. The study was based on inductive reasoning [34], a method well-established in scientific research, as shown by studies to perform content analysis [30, 31]; increase fidelity in the approach through computer-aided qualitative data analysis software [37]; and validate observational research [38]. The primary research approach consisted of a three-phase process incorporating the concept of triangulation. First, a literature search and analysis were performed to clarify the gaps and formulate the topics to be investigated. Next, the actionable research effort included a focus group and one-to-one interviews to gather practitioners' insights. By applying content analysis, the main themes were extracted. Last, a conceptual model was formulated, with four thematic groups of barriers to integration of DFA in MFD. Each step is detailed in the next sub-sections. The three phases are illustrated in Fig. 1, highlighting how they all contributed equally to the study's conclusions. This research applied interpretivism to understand the perspectives and experiences of participants, while also considering pragmatism to ensure practical applicability of the findings.

Table 1 Details of data sources for the mixed-methods approach

| Data source | Content |
| --- | --- |
| Main author's previous publication | Previously published literature review [15] |
| Additional academic material | Literature published in academic journals and conference venues 2022-2024 |
| Focus group | Hybrid (in person + online), two hours focus group with 10 participants asked to reflect on their perspective, guided by input from the focus group mediator (one of the authors) |
| Interviews | 5 semi-structured online one-to one interviews with 5 of the industrial practitioners who participated to the focus group, focused on first-hand experience and personal insights |

[Fig, 1 around here]

## *Data collection*

Secondary data collection involved the analysis of literature realized in a previously published article [15], reinforced by a survey of additional academic readings as presented in the THEORETICAL BACKGROUND section. For the supplementary bibliometric analysis, the Scopus journal database was queried using specific keywords: 'design for assembly', or 'DFA', and 'product architecture'. Selection criteria and a screening of abstracts as in the previous study was performed, to ensure consistency and only relevant sources. Scopus was chosen for its extensive repository of published scientific content. Through a thematic analysis of the literature, the primary gaps in knowledge and practice were identified, which informed the key questions for the investigation [34].

The next step involved supporting the acquired knowledge with primary data collection, drawn from a combination of a focus group and a series of semi-structured interviews. This investigation focused on the main topics identified in the literature, enriched by the authors' expertise and experience in industrial practice.

In June 2024, a focus group was organized in Stockholm with participants attending both in person and via video conference due to their dispersed locations. The attendees were selected for their extensive experience in the Swedish manufacturing sector, established in modularization practices and assembly. The group primarily consisted of skilled practitioners and one accomplished researcher, all with several years of relevant experience. Additionally, a selection of top-performing students from the IPU (Production Engineering) department at KTH, Stockholm, participated. These students had previously excelled in courses MG2020 – Modularisation of Products[2] and MG2040 – Assembly Technology[3], thus holding both strong theoretical knowledge and practical experience. Both courses included hands-on projects that required tackling various aspects of modularization and assembly with real products, ensuring that the students brought valuable insights to the focus group.

Five of the practitioners who participated were also involved in one-to-one interviews. These interviews aimed to extend the knowledge extracted during the group session and were structured to draw from individual experiences and gather more personal and detailed insights. The interviews were semi-structured and lasted around half an hour each. They were conducted online via video conferencing to accommodate busy schedules.

---

[2] MG2020 Modulindelning av produkter: https://www.kth.se/student/kurser/kurs/MG2020
[3] MG2040 Monteringsteknik: https://www.kth.se/student/kurser/kurs/MG2040

A complete overview of attendees to the focus group and one-to-one interviews, and related experience, is in Appendix A. Copies of the focus group structure and interview questions are in Appendix B. The group discussion and all interviews were conducted in English, recorded with participants' signed consent, and transcribed verbatim. Before each activity, participants were briefed on the study's objectives, the intention to publish the findings and the ethical guidelines guiding the research. They were assured that the recordings and any personal information would remain confidential and would not be disclosed, shared, or misused.

### *Data analysis*

Given the necessity of analyzing the combination of focus group data and one-to-one interview data, content analysis was employed as the method for data analysis, a technique equally applicable to both. This is an approach well-suited for systematically interpreting qualitative data from various sources and enable their presentation in summary form, while allowing the identification of patterns, themes, and categories within the data [39]. The unit of analysis for developing the coding system was the whole group while analyzing the group activity, and the individual participants when tackling the interview transcripts. Once established, the coding system was applied systematically across all transcripts.

Thus, a detailed content analysis of these primary data was conducted to examine the narratives and experiences shared by the experts. Through this analysis, recurring themes and specific challenges were identified, as well as opportunities for improvement. The focus on individual experiences and perspectives allowed us to capture a rich, detailed understanding of the practical realities faced in the field.

In the last step the evidence collected in the previous stages were put under further scrutiny through inductive reasoning to formulate a conceptual framework to better understand and structure the challenges and areas of improvement for the MFD method.

## **RESULTS**

The knowledge shared in the Additional literature section illustrated an array of applied and theoretical subjects. To refine it, additional research as described in the METHODOLOGY section was performed.

These activities led to the formulation of topics discussed during the group and individual interviews. In this section, the results of the content analysis process are presented, organized in the seven themes that emerged from the data. These are named as follows and the surrounding conversation is presented in this section. For a summary description of the themes, refer to Table 2. In the text flow, the most relevant citation are presented in 'quotes', so that the reader could better understand the view of the participants to the group and individual interviews and gain a general feeling of the discussion that was performed.

Table 2 Summary description of the main themes, results of the content analysis

| Theme | Description |
| --- | --- |
| *Assembly strategy* | Focuses on the optimization of assembly processes by aligning with strategic decisions, such as when and where to finalize product customization (late differentiation) and whether to relocate assembly activities (delocalization). |
| *Advanced technologies* | Examines the impact of automation, AI, and machine learning in streamlining assembly, improving precision, and reducing assembly complexity. |
| *Collaboration* | Emphasizes the value of cross-functional and inter-organizational collaboration, involving clients, suppliers, and internal departments to ensure smooth information flow and synchronized efforts in the assembly process. |
| *Design* | Covers the creation of effective assembly instructions, design of interfaces, and application of DFX principles to ensure ease of assembly and product durability. |
| *Workforce* | Addresses the influence of technological advancements on workforce engagement, highlighting the importance of operator training and incorporating feedback to improve both process efficiency and job satisfaction. |
| *Information and visuals* | Focuses on the accurate and accessible representation of modular information, utilizing robust documentation and visualization tools to improve understanding and communication among stakeholders. |
| *Performance evaluation* | Involves assessing modular architecture through performance metrics and validation processes, with an emphasis on involving operations teams early to ensure practical and efficient design implementation. |

### *Assembly strategy*

One significant part of the discussion revolved around assembly strategies, a crucial step when deploying a manufacturing plan. Practitioners discussed various strategies such as the hamburger, pyramid, base unit, and Christmas tree approaches to find the optimal sequence. However, they noted that in real-world scenarios, determining it is challenging due to complex supply decisions and production equipment considerations.

> 'So, it helps to understand how the final assembly could be arranged, but there is still a lot of information that will be needed or could be developed […]. If you try to strive for hamburger assembly because that is the obvious, the easiest one (just put the modules on top of each other), but in real life, it's almost impossible'

Additionally, the sequence heavily depends on production volumes and the overall strategy for the module.

Some practitioners highlighted the relevance of DFA guidelines within the assembly sequence, such as considering the orientation of parts (e.g., turning upside down), and number of directions of assembly and the weight and number of operators needed. This indicates a need for integrating DFA principles early in the development to streamline processes and reduce potential inefficiencies.

Strategic decisions for postponing product differentiation were also discussed, as make-or-buy decision affect the supply chain. The participants suggested having modules ready for final assembly based on customer orders to improve time-to-market and customer satisfaction. However, they also pointed out the lack of flexibility in achieving order requirements, especially when dealing with a high number of variants.

> 'Many smaller companies have the strategy to just take orders, and then the flexibility is not very high for the customers to be able to change the construction or change the component in a very late phase, which makes it difficult.'

Late differentiation was identified as beneficial for low-volume, make-to-order production models, although it may negatively impact time-to-market. Decoupling functions and modules to achieve late differentiation was seen as a strategic advantage, particularly when considering geographical dependencies and external factors like pandemics or geopolitical events.

Delocalization, or moving production to different locations, was another significant aspect. Practitioners stressed that delocalization decisions should align with the overall assembly strategy of each module. Furthermore, they highlighted the challenge of losing knowledge and expertise when production is moved away from the main assembly site, which complicates in-house production capabilities.

> 'Assembly, according to my experience, at least in the Western companies in Europe, has been regarded as some sort of necessary evil, and that is the reason why we have bought most assembly from other countries.'

The recent trend of shifting production to multiple continents instead of a single location (e.g., China) was discussed as a response to issues like the pandemic or the current geopolitical situation, with the risk of concentrated production. A more balanced approach might be needed, where high-volume production might be outsourced, but critical or high-quality modules could be kept closer to home for better control and collaboration with trusted suppliers.

The effects of modularization on assembly processes and overall company strategy revealed that standardized interfaces are crucial for achieving high-level modularity, facilitating easier assembly and integration of modules. The importance of formulating a cohesive assembly strategy within and between modules was stressed, noting that each module's strategy influences decisions related to location, storage, automation, and volumes. The biggest impact of MFD is in incorporating strategic information into the information model, which must be easily accessible and comprehensive to support decision-making. This integration aims to create a robust framework that links MFD and DFA, ultimately improving the design and assembly of modular products.

> 'We are capturing the information, and the information carrier is really the modules. But as we are describing the modules, that is a heading and you need to have a strategy, you need to have a certain technical solution, need to have an interface. So that is the core with the modular function deployment.'
>
> 'What is happening with MFD methods compared to other methods […] is that you are starting to define and finalize a concept before you are dealing with the details. And then you also need to discuss assembly topics on the right level.'
>
> 'The basic idea until […] five years ago, was that we start to design and define the modules, that is step number one. And then, when we design each module, we apply the DFA thinking in the detailed design of each module. Very little has been thought of when it comes to the final assembly, how you assemble the modules together. That is still an area where the knowledge could be developed.'

### *Advanced technologies*

The level of automation required depends largely on the ambition and volume levels. Early definition of these parameters is crucial to derive appropriate automation needs. Handling few variants is easier and more cost-effective, while numerous variants necessitate either manual setups or highly flexible and reconfigurable automated systems, which are often expensive. Moreover, it is important to highlight the direct relationship between the increase in number of variants (and consequent increase in sales) and the increase in internal costs.

Concept design for automation was emphasized, particularly the importance of designing and determining feeding system requirements early in the process. Practitioners stressed the need to decide whether to transform existing production setups or integrate new automation systems based on current production requirements. A key strategy to facilitate automation is the reduction of the number of parts, aligning with DFA principles. This reduction not only simplifies assembly processes but also enhances automation efficiency.

> 'I realized suddenly the relation between design and production. So again, I think if we can apply DFA thinking, automatic assembly will be much easier to apply. Again, there will be some things that need to be rethought. In the automatic assembly, for example, feeding all the components and things like that.'

With the goal of implementing the Industry 4.0 enablers, integrating AI and machine learning algorithms presents both opportunities and challenges. The potential for AI to arrange modules and suggest optimal assembly strategies might be very high, though the method of feeding relevant information remains a crucial question, both from the technical perspective and from the regulatory point of view, considering confidentiality issues as well as new protocols. The capture of tacit knowledge through large language models (LLMs) was identified as a significant opportunity, enabling the retention and retrieval of information that might otherwise be lost when employees leave the company. However, despite the increase of interest in AI and LLMs for design, the process is still in its infancy, requiring additional understanding and research effort. Nevertheless, the exploitation of data models from MFD through LLMs was highlighted as a primary objective.

'Currently, we have quite a lot of tools and templates to aid us in that thought process, but it can certainly be more efficiently done if we have a large language model at our disposal. But, on the other hand, it's quite sensitive data, in terms of when it comes to the client's data. There can be some challenges in training an algorithm, some legal challenges in that.'

### *Collaboration*

It is particularly difficulty to connect customer inputs to assembly choices. Including customers in the discussions and decision-making process was suggested as a strategy to ensure their needs are adequately addressed. However, there still is a need to develop quantitative measures for clients to evaluate proposed concepts effectively, as this would enhance the clarity and objectivity of feedback.

'I feel that finding the right balance with what the customer demands and the existing design that we have and the strategy that we want to adapt going ahead [is hard]. We should be able to find the right balance between that and incorporate those demands properly so that we do not end up changing the whole architecture.'

At the same time, collaboration with suppliers is crucial to ensure correct delivery of components and modules to streamline the assembly process, despite possible differences in volumes and strategy. The discussion revolved around suppliers often lacking specific knowledge of the processes, which led to mishandling or misinterpreting information about the modular design. Sharing knowledge and including suppliers in the development loop can bridge this gap. Finding ways to merge the processes and knowledge bases of suppliers and companies is crucial, and clear communication and collaboration can help.

This happens to be especially true when dealing with communication across different departments within a company, and with building correct company culture. The starting point should be establishing cross-functional teams with the identification of all the needed stakeholders. Moreover, teams need to be collaborative, interested in making a change, and well-informed about the process direction. A lack of information often hampers the decision-making processes. Thus, the link to implementation of concurrent engineering practices, which can streamline processes and improve collaboration. However, finding the right timing for decision-making and ensuring strategies are communicated effectively is not a trivial task. Larger companies, in particular, face challenges in spreading company culture and facilitating communication between departments.

> 'So, having a cross functional team, I think it's the best way to balance and manage those different perspectives'
> 
> 'And then sometimes it's just a question of communication.'
> 
> 'A company strategy very often gets stuck in the boardroom. It's not deployed all over the company, so it is not sure that the full strategy is known by the R&D people and the assembly or production departments. So, it will help them include production, so the product really matches the company strategy.

Particularly overlooked seemed to be the incorporation of maintenance considerations. Maintenance personnel need to be involved in the design process from the beginning. Despite resistance from production teams who prefer to contribute only when a design is finalized, early involvement is crucial. Maintenance requirements should be integrated into internal criteria, to ensure they are considered during the design phase.

## *Design*

One surprising outcome emerged during the focus group and interviews: the importance of considering the design of instructions for assembly, and which methods are preferable for presenting instructions. Creating effective instructions for operators is essential, especially given the variety of modular products. Simplifying instructions and tailoring them to the specific needs of operators, including involving them in the process and providing clear signals to avoid major errors, were identified as crucial strategies. Methods to properly design them, current and best practices in information presentation, were mentioned as potential tools to enhance the clarity and effectiveness of instructions. Additionally, when designing modules and assembly instructions, considerations for color blindness and other accessibility issues should be incorporated to ensure that instructions are clear to all users. Methods for presenting instructions to operators could be varied and depending on the designer, on the objective and on the production system. Signaling was identified as the best method to guide operators. Other methods that include augmented reality (AR) and virtual reality (VR) have potential, though they might be too cumbersome for practical use in all scenarios.

> 'The product design should tailor how the product will be assembled, and it should show guides, because actually the instructions are not well used.'

Incorporating DFX principles, which include considerations for upgradeability, disassemblability and recyclability, could also represent a vital point for sustainable product design. Design for sourcing and service were also mentioned as important aspects that should be included in the design criteria. Considering the global footprint into the design strategy was recommended to achieve more sustainable designs, although cost considerations often take precedence. Practitioners noted also that these aspects might often go overlooked during the MFD process, due to the lack of a structured way to document and detail DFX and circularity aspects, suggesting the need for a more formalized approach.

> 'I see that we try to incorporate more and more information over time into the architectural model that the MFD method helps us to create. And I feel that we're lacking a structured way of writing down and detailing the aspects of DFX and circularity and different considerations within assembly.'

The design of interfaces directly impacts the ability to assemble and disassemble products, linking DFX principles to interface design. Specifically, effective interface design between modules is critical for improving assembly and overall integration. The role of DFA concepts was emphasized for standardizing interfaces between modules and enhancing efficiency of the assembly process. While doing that for every module may be unrealistic, at least the method of assembly can be planned to be the same, even if modules might vary in size.

> 'I can say that the principle of having one interface for every module is a utopia. Don't take my word for it, but my impression is that this is quite rarely achieved. You will usually have different variations of interfaces.'

> 'Depending on the project, it can be a consideration whether or not that [interface] requires the same type of assembly method. So, even though you might not be able to standardize the interface, you can at least standardize the method of which you assemble the two modules onto that interface.'

The importance of visualizing interfaces during the creation of the architecture was discussed, and particular importance was given to prioritizing the better solutions. A suggestion was to try and develop a quantitative evaluation to improve design outcomes, however there is often a lack of quantitative data that could be exploited to properly choose the best alternative. Balancing decision making is difficult before establishing a detailed design, and it most likely will result in changes along the way. This should therefore be done iteratively as a continuous improvement process.

### *Workforce*

The increasing levels of technology in assembly work can influence workforce engagement both positively and negatively. One primary challenge is staff turnover, which reduces the accumulated knowledge within companies, leading to a need for simplifying processes to maintain engagement and efficiency. Early involvement of operators in the process is critical for maintaining their engagement and ensuring their insights are incorporated into the design and assembly stages.

> 'And we tried to talk about involvement because even though the experience is changing, we want to involve the operators early in the process so that they are more engaged and can do continuous improvement, for instance.'

However, keeping the workforce happy and engaged presents challenges, as different individuals have varying ideas of what constitutes engaging work. It is beneficial to consider the entire work routine, rather than just focusing on task-specific instructions, to create a more engaging and fulfilling work environment. Creating teams of operators involved in the assembly process of several different modules, or on different steps on the final assembly line could foster engagement, collaboration and competitiveness. Incorporating operator feedback is vital for improving instructions and assembly processes. Operators often express a desire to learn various tasks and become multi-functional, highlighting the importance of providing diverse learning opportunities. Additionally, designing the workplace to suit operators' preferences can enhance their productivity and satisfaction.

Another step is training and developing operators and stakeholders, essential to adapt to advanced technologies and new processes. A significant issue is the perceived lack of competence retention within the workforce, due mainly to the high turnover of people. This requires targeted training programs. Companies often aim to reduce learning time by simplifying processes and creating shorter sequences. However, it is crucial to encourage operators to excel in various tasks and stations to foster a versatile and skilled workforce. The reluctance to invest in training due to fears of losing value if employees leave is another barrier. Much valuable information is tacit and resides within individuals, emphasizing the need for effective tacit knowledge transfer methods.

## *Information and visuals*

The representation and transfer of module information through all design phases is critical. Modularization often fails to properly address representation and visuals, focusing on the primary information of a product – deriving mostly from customer specifications – and the relationships between modules and product properties. Nevertheless, such representations are essential for developing a clear architecture and creating a structure that enhances knowledge transfer and facilitates downstream processes such as detailed design, manufacturing, and assembly.

Understanding how information and visuals are managed in the process can significantly enhance the clarity of product development. Practitioners emphasized that clear documentation and visualization tools are essential for consistency and collaboration across different teams and projects. Robust documentation practices, the integration of advanced visualization tools, and the need for improved knowledge transfer mechanisms are ideal. Practitioners unanimously agreed on the necessity of detailed and accessible documentation to maintain consistency and clarity in design and assembly processes.

> 'Documentation improves [the process] and makes it easier to get out the information again if you need them at later stages.'

While documentation is the primary method currently used, there is a clear need for more advanced and integrative tools. Sketching is valued for its simplicity and effectiveness in the early design stages. Color coding and matrix structures were suggested as additional methods to organize and present information effectively.

> 'I think that's still under development but making it visual may be the challenging part. It's always easier if you get the sketch or something, to get what you mean, instead of looking into a matrix.'

Practitioners expressed a desire for consolidated visualization tools that could integrate all relevant information and visuals in one place, enhancing clarity and accessibility, thus making collaboration and decision-making easier. CAD models and diagrams were highlighted as essential for illustrating the product portfolio and aiding in the visualization of complex assemblies.

Furthermore, the integration of Product Lifecycle Management (PLM) systems was suggested to systematically record all decisions, aiding traceability and facilitating easier project handovers. While reusing information can save time and resources, it must be managed carefully to protect sensitive data. Despite the available tools, the reliance on traditional methods such as Excel and PowerPoint was seen as a limitation, as these tools do not provide the interactivity and integration required.

> 'A visualization of the information model and the impact, most of all the impact, that different types of decisions have on the information model at large would be really beneficial to have, if it were to come as an addition from the manufacturing and supply chain perspective.'

### *Performance evaluation*

Validating the architecture of modular design is crucial for ensuring that all components work together seamlessly. Customers appreciate having a quantitative validation of the architecture, which poses a challenge in terms of identifying what can be quantified. A preliminary visualization of components included in a module and relative interfaces is valuable for this validation process. However, some companies experience what is known as 'overdocumentation' and find it difficult to capture the value of the amount of data, as it feels cumbersome in relation to the benefits it brings. This threshold is subjective and varies greatly from one customer to another, more often than not resulting in gaining new, important insights through the process of documenting.

Evaluating interfaces based on criteria like electrical connections helps identify good and bad interfaces. Consequently, an evaluation matrix, which assesses single interfaces between modules, can help determine the most efficient design by considering factors like the number of directions of assembly and the number of operators needed. Involving operation experts early in the design process and capturing their knowledge can help address the difficulty of communicating design details to them.

'When it comes to manufacturing and operations, that's where there is usually some sort of gap in trying to communicate the power of the architecture, because it's not designed fully in detail.'

To streamline the assessment and validation of the architecture and help communicate its significance, various methods might be employed, such as linking DFA evaluations to the impact of modularization choices. However, it was mentioned that it would be essential to start with the process of cleaning out unnecessary solutions. Then, the development of a specific DFA index for modules would give a number to the different decisions, though this appears to be particularly challenging.

'The first [thing] is to clean out what is the concept you really need and then you can apply [DFA] and I think that is where we should [ask]: "OK what is the DFA index of the selected solution?", and maybe you weigh with that; you first select out of ten different technical solutions, remove those [unnecessary], and do the DFA index analysis.'

Establishing clear evaluation criteria remains difficult but selecting and collecting a list of best practices and libraries of knowledge about assembly methods, interfaces connections and joining techniques could greatly contribute to the process. Maintaining a history of tools, production setups, and methodologies can inform designers and facilitate the reuse of proven solutions.

'Different people are motivated by different things and usually we try to convey it through some sort of KPI measurement. […] we have achieved a certain degree of part number reduction, or we can see this amount in increased sales, or we have decreased this current cost of complexity and so on, in order to motivate different stakeholders within the company.'

Defining Key Performance Indicators (KPIs) is essential for tracking and measuring performance effectively. Performance measurements are essential for evaluating the practical aspects of modular designs. Key considerations include the weight of components, the number of interfaces attached at the same time, and the direction of assembly. Reducing the number of parts without compromising the design is a paramount criterion, but evaluating this in the early design stages is challenging. Side-by-side comparison of designs facilitates discussion and evaluation. Cost considerations, such as the cost of change and cost-efficiency, are nevertheless considered crucial, and often are the sole critical factor leading the decision-making process. Guidelines and checklists created by cross-functional teams can aid in standardizing evaluation criteria. However, internal criteria, which are usually defined through group discussions, need to be better connected to the assembly processes and complexity of assembly. This connection is often loosely defined, making it difficult to prioritize criteria effectively.

### *Conceptual model development of the barriers*

Practitioners widely recognize the development of DFA integrated modular architecture as a key goal in the field. However, achieving widespread adoption of these concepts remains challenging. Despite significant interest and ongoing research efforts, several obstacles block the formulation of a structured approach, hindering broader acceptance.

A conceptual model was developed, based on the synthesis of literature and practitioners' insights. The limitations identified through content analysis can be further grouped into four main thematic categories: Technological, Economic, Regulatory, and Organizational (TERO). This aims to simplify the outcomes by grouping common themes and related barriers that emerged. The conceptual model is schematically shown in Fig. 2. The four thematic groupings are detailed below.

(1) Technological (T) barriers encompass issues related to the current state of technology and the most recent developments. They include challenges such as difficulties in integrating new technologies with existing processes, and limitations in technology's capability to handle complex design requirements, such as incorporating recyclability and addressing hazardous materials.

(2) Economic (E) barriers include financial constraints and cost-related challenges associated with implementing a structured approach. They encompass issues related to the adoption of expensive technologies, financial risks associated with transitioning to new methods, and concerns about return on investment. Economic barriers often reflect the broader economic impact on decision-making in modular architecture.

(3) Regulatory (R) barriers consist of challenges related to compliance with existing or upcoming regulations and standards. They can delay the adoption of new methods if these are not aligned with current standards or if they introduce additional compliance costs, such as those associated with new AI regulations. Other related issues include data management, adherence to security protocols, and the problem of confidentiality when handling proprietary and sensitive data from manufacturing firms.

(4) Organizational (O) barriers refer to obstacles within an organization, such as resistance to change, inadequate staff training, and lack of alignment between different departments or stakeholders involved in the process. Here also the aspect of sharing information needs to be mentioned, including the limitations of CPQ/PLM/ERP systems integration capabilities. Organizational culture, structure and the size of the company can also play a role in defining the magnitude of such barriers.

Each of these thematic groupings highlights a different aspect of the barriers that need to be addressed to construct a structured workflow for their consistent industrial implementation. The barriers identified here will be explored in more detail in the DISCUSSION section.

[Fig. 2 around here]

## DISCUSSION

We analyzed recent developments in the manufacturing and design industry, focusing on the integration of modular architecture and assembly processes. By employing a content analysis approach, we aimed to understand the underlying themes from the perspective of industrial practitioners working closely with MFD method, applied to different industries. Our findings revealed a strong push towards sustainable and flexible manufacturing that can be achieved through the adoption of modular designs and will most likely acquire even more leverage through the integration of design for assembly concepts. This vision is supported by the extensive experience of the specialists, who are continuously striving for innovation to improve production.

The discussion from the group sessions and individual interviews, coupled with the recent surge in publications in AI-methods for product design, suggests that interest in this area will continue to grow significantly. This is largely attributed to advancements in LLMs, enabling more sophisticated data analysis, enhancing creativity, and supporting the automation of complex design tasks. LLMs offer advanced capabilities in natural language understanding and generation, facilitating better communication of design concepts and requirements. They also help in generating innovative design solutions from extensive datasets that include historical design data, market trends, and customer feedback. This capability to learn from diverse sources and generate contextually relevant suggestions is pushing the boundaries of traditional product design methodologies. Moreover, the integration of AI and LLMs in product design is a significant shift towards more data-driven and user-centric design approaches. Designers can leverage AI to predict customer preferences, optimize product features, and reduce time-to-market. As AI tools become more accessible and user-friendly, their adoption in product design is likely to increase, driving further innovation and efficiency.

Despite these promising advancements, several gaps remain in developing a structured approach to address assembly issues during the creation of a modular architecture. Through our results, different aspects and challenges came to the surface. Within each thematic area, the TERO barriers emerged.

One main theme is the recurring emphasis on the need for standardized metrics and clear guidelines to impact decision making from multi-disciplinary groups in the early stages. Standardizing interfaces and including structured evaluation criteria is seen as crucial for successful strategic modularization aimed at ease of assembly. The lack of a unified workflow hampers the efficient formulation of modular architecture.

Another subject is the necessity of identifying and involving relevant stakeholders early in the process, including clients, suppliers, and cross-department teams. This includes production operators, as well as maintenance employees, where appropriate. This inclusive approach ensures that diverse perspectives and expertise are integrated, leading to a robust modular design, striving for optimization of functionality, manufacturability and even recyclability, reusability, remanufacturability, through the product life cycle. This idea is tied into the need for effective communication and knowledge transfer mechanisms, which are critical for capturing and leveraging tacit knowledge, regarded as a problematic area for companies with high turnover of employees. Nevertheless, this is often seen as a very costly endeavor for companies, since it is not praxis for production operators, suppliers, and so on, to be present at this stage: they need to be kept motivated in participating and engaged in the process.

From a technological and economic standpoint, the main barriers center on current limitations of existing systems and the necessary advancements to integrate modular designs effectively. The lack of standardization in components and interfaces hinders the integration of modular parts, while the rapid pace of technological change creates uncertainty and risk for manufacturers, considering the big investments needed, for example, to deploy automatic or autonomous systems. Additionally, the need for advanced software and tools to manage and optimize modular manufacturing systems poses a significant challenge, particularly for smaller firms with limited resources. The long-term financial benefits, such as reduced lead times and increased flexibility, may not be immediately quantifiable, making it difficult to justify the expenditure to stakeholders. Aligning economic incentives across the supply chain is difficult, as different actors may have conflicting interests and priorities.

Regulatory barriers include the legal and compliance aspects that impact the adoption of modular architecture. Manufacturers need to be aware of existing regulations that stand behind the innovative approaches currently emerging for optimal modular design. The regulatory environment can also vary significantly across different regions, adding an additional layer of complexity for global manufacturers, for example the EU AI Act: first regulation on artificial intelligence[4] only applies to EU countries. The significance of data-driven decision-making is particularly relevant in this context. The EU's evolving regulations underline the necessity for transparent, accountable, and data-driven approaches to ensure compliance and mitigate risks. However, one of the main challenges at this early stage of modular architecture development is the unavailability of sufficient data. This lack of data is one of the reasons behind the necessity of gathering multidisciplinary groups to ensure that decisions are well-rounded and consider diverse perspectives. Engaging experts from legal, technical, economic, and organizational backgrounds can help navigate these complexities and develop robust, compliant, and sustainable modular solutions.

Organizational barriers mainly involve the internal dynamics and cultural aspects of firms. Resistance to change is a common theme, as employees and management may be reluctant to adopt new practices and technologies that disrupt established workflows. The shift towards modular architecture requires a transformation in organizational structure and processes, which can be met with significant inertia. Furthermore, the need for cross-functional collaboration and communication is heightened, necessitating a cultural shift towards more integrated and flexible working practices. Communication gaps and siloed knowledge within organizations inhibit collaboration and the integration of diverse expertise.

### *Limitations and future work*

While the benefits of employing interpretivism, such as gaining a deeper understanding of complex social phenomena and promoting discussion, are well-documented, this approach has notable limitations in the context of our study.

---

[4] EU AI Act : https://www.europarl.europa.eu/topics/en/article/20230601STO93804/eu-ai-act-first-regulation-on-artificial-intelligence

These limitations include the following.

(1) Limited generalizability, as the findings focus on specific context and social constructs, such as the key role of modularity in achieving successful outcomes in mass customization. This is relevant for our study, as the barriers identified may be unique to the expertise and context of the practitioners involved.
(2) Researchers' bias, as the data was interpreted within our research group, employing our perspectives and biases, which can affect the reliability of the findings. Despite following strict protocols and an accurate methodology, this may influence how barriers are identified and analyzed, potentially skewing the results.
(3) Review limitations, as the process of reviewing and conducting thematic analysis may be constrained by incomplete search practices and potential transcription errors. This can affect the accuracy and comprehensiveness of the thematic clusters and groupings.
(4) Reliability of the findings, which may be impacted by the inherent subjectivity in interpretivist and pragmatic approaches.

Despite these limitations, the interpretivist approach has provided valuable insights and generated new aspects that can inform future research. For instance, the conceptual model for identifying barriers across four key thematic categories—technological, organizational, economic, and regulatory—serves as a starting point for advancing knowledge in this area. Future work should focus on developing targeted strategies to address these barriers.

Future research could prioritize the development of advanced software and visualization tools that support real-time data sharing and assembly simulation, addressing current data availability limitations. Additionally, standardizing evaluation metrics and criteria specific to DFA within modular product design will be crucial for enabling consistent benchmarking and refinement. Enhanced collaboration and connection between researchers and industry practitioners, particularly in the areas of product design, assembly, and modular architecture, are also essential. This collaboration should include cross-functional training initiatives that help overcome organizational resistance and foster a culture of interdisciplinary integration in decision-making.

This work, alongside the previously published literature review, forms part of a larger research project. Together, these papers aim to provide a comprehensive overview of the current situation around integrating DFA with MFD, synthesizing state-of-the-art academic findings with insights and experience from industry practitioners. This foundation is intended to guide future efforts in creating and evaluating supports to improve design practices based on DFA and MFD. Integrating technological know-how with practical industry experience is vital to creating effective, adaptable solutions advancing the field. The next studies of this research group will focus on understanding the most relevant flexible practices to meet evolving regulatory requirements, as well as creating a structured support to include assembly considerations in MFD that addresses at least some of the barriers highlighted in this research and can be easily evaluated by industry.

## CONCLUSIONS

The outcomes of this study reveal some barriers to integrating assembly concepts within MFD. Across all thematic groupings—technological, organizational, economic, and regulatory—there is a clear need for a coordinated and collaborative approach to overcoming these obstacles. Technological advancements must be supported by organizational readiness, economic viability, and a supportive regulatory framework, showing the importance of stakeholder engagement and the need for industry-wide initiatives to drive the adoption of modular architecture.

Discussions from focus groups and interviews indicate a shared belief in the potential of modular design to transform manufacturing processes despite the challenges. There is a collective understanding that addressing these barriers requires not only technological innovation but also a significant shift in organizational focus, essential for realizing the benefits of modular architecture and achieving a more efficient and sustainable manufacturing future.

While integrating DFA in MFD presents numerous challenges, the potential benefits make this effort highly worthwhile. The primary obstacles call for the development of targeted supports (tools, guidelines, indices, including metrics that would help companies benchmark their progress and identify best practices) derived from cross-industry collaboration and backed by all stakeholders. Moreover, to build on the presented findings, future research should focus on validating the proposed conceptual model in diverse industrial settings to assess its effectiveness across various sectors and production environments. This validation process could provide valuable feedback to refine the model and enhance its impact on industry practitioners. What was shown includes also a significant potential in exploring AI-driven tools and advanced software solutions. Leveraging machine learning and data analytics could enable more accurate assembly simulations, optimize modular design processes, and improve decision-making during the early stages of product development.

In conclusion, this study provides a structured overview for understanding and addressing the barriers to DFA integration in modular architectures. By offering a conceptual model and identifying key areas for improvement, it contributes to the broader methodology of modular product design. As the industry continues to adopt more modular approaches, the insights from this research can guide practitioners in creating more cost-effective, easy to assemble, and high-quality modular products, ultimately supporting the shift toward a more sustainable manufacturing landscape.

## ACKNOWLEDGMENTS

The authors would like to express their most sincere gratitude to all the participants of the focus group and interviews, practitioners and students alike. The authors are also deeply grateful to Modular Management AB[5] for graciously hosting the focus group at their offices in Stockholm.

## DECLARATION OF INTEREST STATEMENT

The authors report no conflict of interest.

## APPENDICES

### *Appendix A*

Table 3 Details of interviewees and participants of focus group ('Exp. years' refers to the numbers of years of working experience in product design or similar activities)

| Position | Job title | Exp. years | Focus group (Y/N) | Interview (Y/N) | Interview date |
|---|---|---|---|---|---|
| Practitioner | Vice president | 16+ | Y | N | - |
| Practitioner | Senior manager | 16+ | Y | Y | 2024-06-18 |

---

[5] Modular Management AB: https://www.modularmanagement.com/en/

| | | | | | |
|---|---|---|---|---|---|
| Practitioner | Senior consultant | 11-15 | Y | Y | 2024-06-14 |
| Practitioner | Consultant | 3-5 | Y | Y | 2024-06-12 |
| Practitioner | Retired | 16+ | Y | Y | 2024-06-18 |
| Academician | Senior researcher | 3-5 | Y | Y | 2024-06-18 |
| Student | - | 3-5 | Y | N | - |
| Student | - | 0-2 | Y | N | - |
| Student | - | 0-2 | Y | N | - |
| Student | - | 0-2 | Y | N | - |

## *Appendix B*

*Focus group structure*

- Welcome and introduction, briefing (purpose, ground rules, confidentiality)
- Discussion topics

(1) Challenges in selecting assembly concepts and assembly strategies in MFD

(2) Organizational or technical barriers when considering assembly early in MFD

(3) Which concepts of assembly are mostly considered early, what are the benefits

(4) How does MFD/how does assembly impact company strategies?

(5) How is modularity represented with MFD, benefits and drawbacks.

(6) How can information be extracted from a modular design later in the process?

- Wrap-up and final thoughts
- Summary and conclusions

*Interview structure*

- Welcome and introduction, explanation of the structure, purpose and confidentiality of data collection
- Main interview questions (all focused on personal knowledge and experience)

(1) Challenges in including assembly in MFD

(2) Data information model knowledge management through and across projects, representation and communication of knowledge

(3) Visualization of information created

(4) Management and creation of interfaces between modules

(5) Uses of AI in MFD or previous projects

(6) Incorporation of DFX considerations

- Final thoughts and potential additional remarks
- Conclusion, summary and reflection on key points

# FIGURES

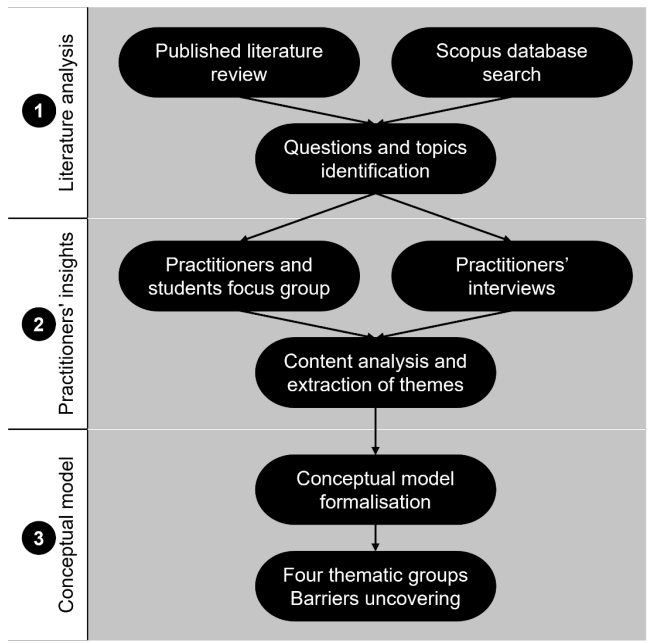

Fig. 1 The three phases of the research process

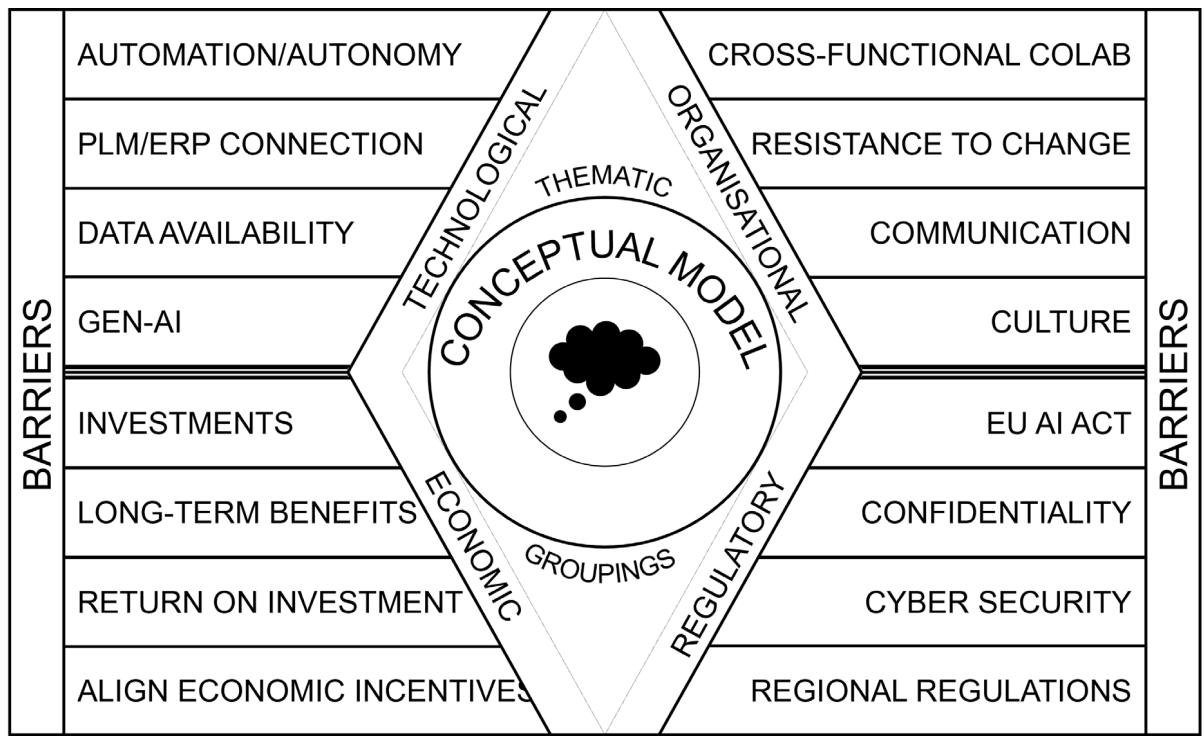

Fig. 2 Conceptual model of the barriers to adoption

# LIST OF FIGURES

# LIST OF TABLES